# A molecular simulation analysis of producing monatomic carbon chains by stretching ultranarrow graphene nanoribbons

Zenan Qi[1], Fengpeng Zhao[1], Xiaozhou Zhou[1], Zehui Sun[1], Harold S. Park[2] and Hengan Wu[1],*

[1] Department of Modern Mechanics, CAS Key Laboratory of Materials Behavior and Design, University of Science and Technology of China, Hefei, Anhui 230027, P. R. China

[2] Department of Mechanical Engineering, Boston University, Boston, MA 02215

*Corresponding author, E-mail address: wuha@ustc.edu.cn (H.A. Wu)

**Abstract**  Atomistic simulations were utilized to develop fundamental insights regarding the elongation process starting from ultranarrow graphene nanoribbons (GNR), and resulting in monatomic carbon chains (MACCs). There are three key findings. First, we demonstrate that complete, elongated and stable MACCs with fracture strains exceeding 100% can be formed from both ultranarrow armchair and zigzag GNRs. Second, we demonstrate that the deformation processes leading to the MACCs have strong chirality dependence. Specifically, armchair GNR first form DNA-like chains, then develop into monatomic chains by passing through an intermediate configuration in which monatomic chain sections are separated by two-atom attachments. In contrast, zigzag GNR form rope-ladder-like chains through a process in which the carbon hexagons are first elongated into rectangles; these rectangles eventually coalesce into monatomic chains through a novel triangle-pentagon deformation structure under further tensile deformation. Finally, we show that the width of GNRs plays an important role in the formation of MACCs, and that the ultranarrow GNRs facilitate the formation of full MACCs. It shows the possibility due to the experimentally demonstrated feasibility to use narrow GNR to fabricate novel nanoelectronic components based upon monatomic chains of carbon atoms.

## 1. Introduction

With the ongoing development of both micro and nanoelectromechanical systems (M/NEMS) and the resulting miniaturization of electronic devices, there has recently been extensive interest in studying the properties of molecular and atomic scale components. For example, devices where electrons hop onto, and off from, a single atom between two contacts would be the ultimate limit. Because of this, creating reliable and stable single-atom chains has long become the focus of research in this field. Experimental studies where a monatomic chain of several Au atoms (Au atom chain of single-atom width) pulled out were first reported in 1998 [1, 2]. In 2003, long linear carbon chains inserted in multiwalled carbon nanotubes were observed [3, 4].

However, interest in the field has shifted dramatically with the recent discovery of graphene, the world's only known 2D material [5, 6]. In particular, because of its remarkable conductivity and other exceptional physical, chemical and electronic properties including nanometric dimensions, quantized conductance, high modulus of elasticity, unusual optical features and electromagnetic response [5-10], graphene is a promising candidate to be fabricated as a M/NEMS component.

Recently, graphene nanoribbons (GNR) whose width is smaller than about 20 nm have been derived both by chemical synthesis and stable narrow GNR without hydrogenation can be obtained by irradiation methods [11-13]. GNRs are of particular interest due to the fact that graphene has been found to become a semiconductor once the width of the GNR becomes smaller than about 20 nm [14, 15]. With regards to the present work, GNR are relevant because MACCs can easily be produced by the elongation of narrow GNR, as shown in previous works [16, 17]. Furthermore, single atomic chains containing more than ten carbon atoms have been observed to have a lifetime spanning hundreds of seconds [11]; in contrast, metallic atom chains have been experimentally observed to become unstable after time scales on the order of seconds [1]. Recent experimental studies have also shown that MACCs derived from graphene have exceptional promise as a conductor due to their outstanding stability and electrical properties[18-20] .

Previous atomistic simulations have shown that MACCs can be obtained by pulling GNR with graphene leads [16]; however, these studies did not analyze or elucidate the mechanisms underlying the elongation process. Furthermore, the critical effect of GNR chirality was also not considered. Finally, these studies did not demonstrate the importance of using the narrowest GNR to achieve very stable MACCs with high fracture strains exceeding 100% under tensile loading.

Therefore, it is the objective of the present work to, using molecular dynamics (MD), molecular mechanics (MM) and density functional theory (DFT) simulations, analyze and shed insights into the factors controlling the elongation process of ultranarrow (i.e. single unit cell wide) GNR. We first demonstrate that highly elongated and stable MACCs are possible from both ultranarrow armchair and zigzag nanoribbons. We then demonstrate that chirality plays a critical role in the resulting elongation process, where ultranarrow armchair and zigzag GNR exhibit a significantly different mechanical response due to the operant deformation mechanisms under tensile loading. Finally, the results suggest that the formation of complete and elongated MACCs is greatly facilitated by being pulled from ultranarrow GNR; this critical point was not observed by previous GNR deformation studies [21, 22] These results should be of technological relevance due to the extensive interest in using carbon-based materials, including MACCs, as the building blocks for novel nanoelectronics and M/NEMS [15].

## 2. Model and Methodology

A schematic of the two different GNRs considered in the present work is shown in Figure 1; snapshots for this and all subsequent figures were taken using VMD, a visualization tool to view MD simulation results[23]. In our work, the MD simulations were performed using LAMMPS [24], while the carbon-carbon interactions were modeled using the adaptive intermolecular reactive empirical bond order (AIREBO) potential [25]. The AIREBO potential has been shown to accurately capture the bond-bond interaction between carbon atoms as well as bond breaking and bond re-forming. The cutoff distance for the REBO part of the potential was set within LAMMPS to be 2 Angstroms to avoid spuriously high bond forces during the fracture process, as described by Shenderova et al. [26]. The time step was chosen as 1 femtosecond, and the simulations were carried out at room temperature (300K) in an NVT ensemble. The GNRs were stretched using tensile loading by applying a ramp displacement that went from zero in the middle of the GNR to a maximum value at the two axial ends of the GNR. The strain rate in the simulations was set as $10^8 s^{-1}$, or equivalently 0.5m/s; we note that this is much slower than the upper bound speed of 30m/s found by Wang et al for MACC formation [16, 19]. The MD and MM simulations were utilized to examine the details of the tensile evolution process, while DFT calculations were utilized to examine the energetic stability of the various GNR and MACC configurations.

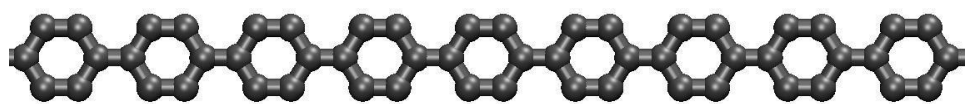

Figure 1-(a)

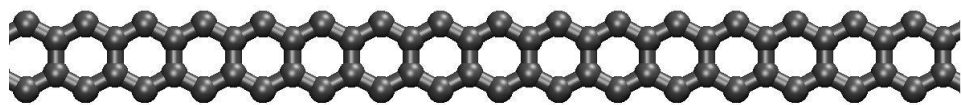

Figure 1-(b)

Figure 1 Schematic of narrowest possible GNR for (a) Armchair GNR (AGNR), (b) Zigzag GNR (ZGNR)

## 3. Simulation details and results

We first establish that complete MACC formation is possible from both ultranarrow AGNR and ZGNR; we will examine the atomistic details of the chirality-dependent MACC formation later. The ultranarrow AGNR we considered in Figure 1(a) had a length of 54.954 Angstroms, with periodic boundary conditions (PBC) in the length direction to mimic an infinitely long AGNR, and had free surfaces in the transverse directions such that finite width effects could be considered. A conjugate gradient (CG)-based energy minimization was first carried out at 0K before dynamic, tensile loading was applied at room temperature as described above; snapshots of the AGNR during the elongation process are seen in Figure 2.

During the initial equilibration, the AGNR demonstrated a rotation and distortion of the initially planar structure, as some of the hexagons have rotated out of their initial planes, as seen in Figure 2(a) and 2(b); this rotation and distortion occurs due to the edge stress induced warping as discussed by Shenoy et al [27]. Specifically, the carbon-carbon bond connecting adjacent hexagons acts as an axis, about which each six-atom hexagon ring rotates. Some of the adjacent hexagon rings are observed to be nearly perpendicular, and the entire AGNR assumes a DNA-like chain.

After a tensile strain of 20%, the hexagon unit cells became significantly deformed, as a noticeable elongation of the six-atom hexagons was observed. By a strain of 35%, MACCs with multiple atom attachments were observed as in Figure 2(c); these MACCS then continued to elongate under further tensile loading. The multiple atom attachments existed due to the inability to completely and instantaneously pull all atoms of a six-atom hexagon ring into the MACC; the last (highly distorted) six-atom hexagon persisted until a strain of 85.5%. Furthermore, at a strain of 133%, all multiple atom attachments were absorbed into the MACC, thus forming a long and complete MACC as seen in Figure 2(d). The average distance between neighboring atoms was 1.64 Angstroms, with a range from 1.46 to 1.72 Angstroms. Under continuing stretching, the bond lengths increased and finally fracture of the MACC was observed at a strain of 150% as in Figure 2(e).

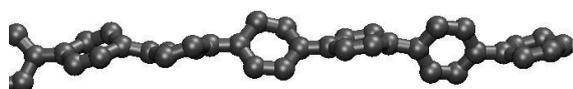

Figure 2-(a)

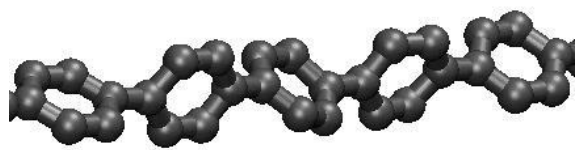

Figure 2-(b)

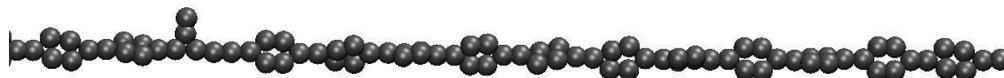

Figure 2-(c)

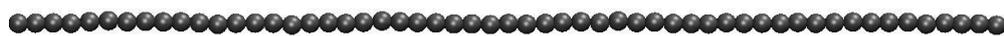

Figure 2-(d)

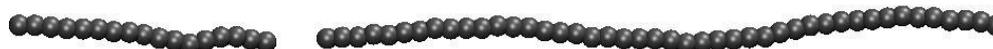

Figure 2-(e)

Figure 2 (a) DNA-like AGNR during relaxation, (b) Another angle of the DNA-like AGNR, (c) The appearance of MACCs with multiple atom attachments at a strain of 35%, (d) Complete transformation of the AGNR into a MACC at a strain of 133%, (e) Fracture of the MACC at a strain of 150%

MM simulations of the same AGNR model were also performed by CG energy minimization. In these MM simulations, tensile strain was applied by fixing one end of the AGNR, prescribing a displacement of 0.05 Angstroms at the other end, allowing the interior atoms to relax to energy minimizing configurations, and repeating. Note that due to the fact that the MM simulations were performed at 0K, the AGNR showed little of the rippling that was observed in Figure 2 at room temperature. A similar tensile evolution process was observed for the MM simulations as compared to MD, including alternating vertical units due to rotation and buckling of adjacent hexagons, reconstruction to MACCs and fracture. More importantly, the MM simulations revealed that energy minimization was responsible for the twisted configuration seen in Figure 2(a) and 2(b). Specifically, the DNA-like structure in Figure 2(a) had a potential energy that was about 0.68% lower than the initial, undeformed configuration seen in Figure 1(a), i.e. the twisted configuration enables the AGNR to relieve the stresses that result from the armchair edges [27].

We studied the energetics of the MACC configurations using both DFT and MM simulations.

In the DFT calculations, an infinite MACC model was established with PBC in all three directions, though in the two transverse directions, the periodic distances were set much larger than 10 Angstroms to ensure no interactions between MACCs. We utilized the generalized gradient approximation (GGA) for exchange correlation functional [28] in the CASTEP code [29], where the energy cutoff was 400 eV and the k-point mesh was set to 1x1x21 to ensure more than 10 k-points in the length direction. After optimization, it was found through the DFT calculations that the MACC indeed existed in a stable configuration, with alternative bonds with lengths of 1.261 and 1.301 Angstroms, respectively, were found. The results are comparable to those of the previous MD simulations after fracture where an average bond length of 1.341 Angstroms was found.

In the MM simulation, we studied a MACC with 100 atoms. The model was considered without PBCs in the axial direction, so we consider bond lengths in the middle of the MACC only to avoid any edge or surface effects. After applying an initial random velocity to generate a representative finite temperature profile, the CG method was applied to find the minimum energy configuration of the MACC. Before energy minimization, three bond lengths were dominant; those bond lengths were 1.12, 1.42 and 1.62 Angstroms. However, after energy minimization, all the bond lengths were found to have a double bond length of 1.332 Angstroms, which is very close to the average DFT bond length of 1.281 Angstroms and MD bond length of 1.341 Angstroms. The results also justify the usage of the classical AIREBO potential for studying MACCs.

To study the effect of chirality on the deformation processes leading to MACCs, we also studied the tensile deformation of ZGNR. As seen in Figure 1(b), the connection between adjacent six-atom hexagons for ZGNR is accomplished using two carbon atoms, as compared to just one atom for the AGNR; therefore, it is expected that the rotation and twist that was observed for the AGNR in Figure 2(a) may not be observed for ZGNR.

The atomistic model used for the ZGNR was similar to the AGNR of 48.944 Angstroms length, with PBC applied along the axial direction, while free surfaces were allowed along the transverse directions to examine the effects of finite width. The ZGNR was first relaxed by energy minimization, then an MD simulation was performed using the same linear ramp displacement profile as was used to apply tensile deformation to the AGNR; the deformation process for the ZGNR is shown in Figure 3.

Interestingly, during relaxation, the ZGNR acted as a suspended rope ladder as shown in Figure 3(a), which again results from relieving the edge stress that results from the zigzag edges [27]. While the entire ZGNR wrinkled significantly, the hexagon rings showed little rotation and instead formed a parallel rope-ladder-like ribbon in which atoms at the edges of the cross direction had nearly the same height, though when thermal fluctuations are discarded in the MM energy minimization simulations, the minimum energy configuration of Figure 3(a) is one in which the parallel edges of the hexagon remain planar, as seen in Figure 3(b).

As can be seen in Figure 3, the processes and mechanisms enabling the eventual formation of a MACC from the ZGNR are completely different than those seen in Figure 2 for the AGNR. In particular, the major deformation process was a reorientation and elongation of the six-atom hexagons initially seen in Figure 3(a) to six-atom rectangles as seen in Figure 3(c). Furthermore, it could be seen that the atoms at the edges form an angle of nearly 180 degrees, as compared to the 120 degree angle observed for the initial, undeformed hexagons. In fact, when all of the hexagons had transformed into rectangles, as in Figure 3(c), the original ZGNR essentially became two parallel lines of atoms.

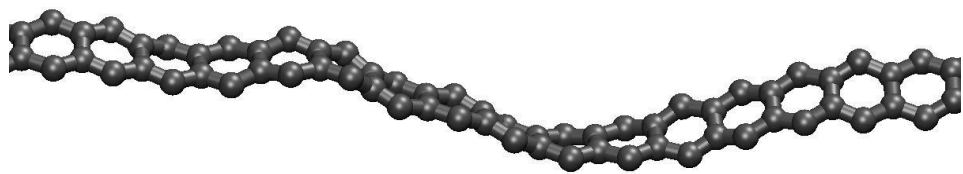

Figure 3-(a)

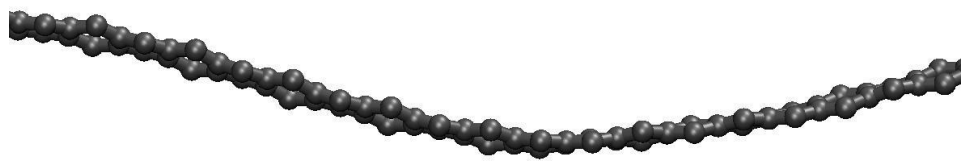

Figure 3-(b)

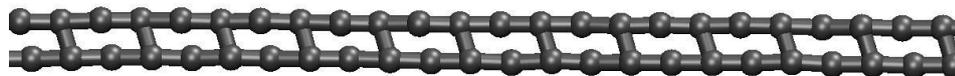

Figure 3-(c)

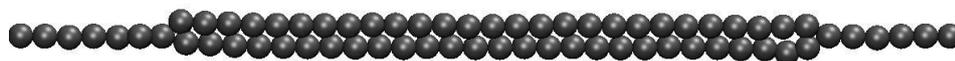

Figure 3-(d)

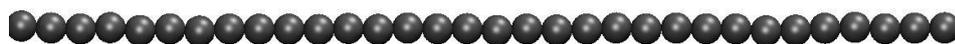

Figure 3-(e)

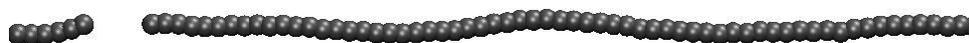

Figure 3-(f)

Figure 3 (a) Rope-ladder-like ZGNR during relaxation, (b) Elimination of thermally-induced fluctuations after energy minimization, (c) Hexagonal to rectangular deformation at a strain of 42.5%, (d) Multiple MACCs at strain of 77%, (e) Formation of complete MACC from ZGNR at strain of 180%, (f) Fracture of the MACC at strain of 188.5%

With further elongation, the two parallel rows of atoms coalesced into a single row of atoms after a tensile strain of 42.5%, as observed in Figure 3(d), where the MACCs surrounded the two parallel rows of atoms in the center of the ZGNR. We note the similarity of this structure to the multiple atom attachments that were observed during the tensile deformation of AGNR in Figure 2(c), which again results from the inability to completely and instantaneously pull all atoms that comprise the rectangular structure seen in Figure 3(c) into the MACC. Finally, at a strain of 180%, the ZGNR had completely become a MACC, and the MACC fractured eventually at a strain of 188.5%. The results in Figure 2(d) and 3(e) suggest that very long and stable MACCs can be derived under tensile loading from both ultranarrow AGNR and ZGNR.

| Chirality | Original length (Angstroms) | Strain A | Strain B | Averaged Bond length at B (Angstroms) | Strain C | Averaged Bond length at C (Angstroms) |
|---|---|---|---|---|---|---|
| AGNR | 54.954 | 35% | 133% | 1.642 | 150% | 1.761 |
| ZGNR | 48.944 | 42% | 180% | 1.712 | 188% | 1.764 |

Table 1 Data comparison of critical strains during the tensile elongation of AGNR and ZGNR. Strain A is the strain when the MACC was first observed, Strain B is the strain when the entire GNR formed a MACC, and Strain C is the fracture strain.

Table 1 summarizes the strain values between AGNR and ZGNR for configurations of interest. Specifically, it shows that MACCs are first seen at different strains for the AGNR and ZGNR, i.e. 35% for the AGNR as compared to 42% for the ZGNR; both the AGNR and ZGNR exhibit large fracture strains of 150% and 188%, respectively. In our simulations, the averaged and maximum bond lengths just before fracture are 1.761 and 1.781 Angstroms for AGNR, 1.712 and 1.790 Angstroms for ZGNR.

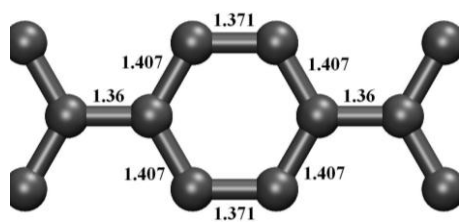
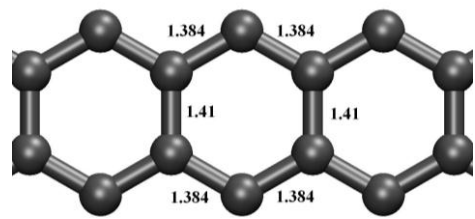

Figure 4-(a)     Figure 4-(b)

Figure 4 Bond lengths of (a) AGNR and (b) ZGNR after energy minimization (Angstrom)

We now examine in detail the specific atomistic deformation mechanisms that enable both the AGNR and ZGNR to form MACCs under tensile loading. Importantly, we will show how chirality effects have a first order effect on the observed deformation mechanisms leading to the MACCs for both AGNR and ZGNR. We first briefly discuss the bond lengths in the AGNR as shown in Figure 4(a); this is done because it is known for covalently bonded materials that the longer covalent bonds tend to be weaker.

A detailed analysis of the deformation mechanism starting with an undeformed AGNR and ending up with a MACC is shown in Figure 5(a-c). There, it could clearly be observed that in going from the six-atom hexagon ring in Figure 5(a) to the intermediate two-atom configuration in Figure

5(b), failure initiated with the two weaker 1.407 Angstrom bonds in Figure 4(a), which were pulled first into the MACC. It is also due to these weaker 1.407 Angstrom bonds that the AGNR initiates the MACC formation at a smaller strain level (35%) as compared to ZGNR (42%), as shown in Table 1. Subsequent to the incorporation of the two long 1.407 Angstrom bonds in the AGNR, the remainder of the two-atom configuration in Figure 5(b) was pulled into the MACC under continued tensile loading.

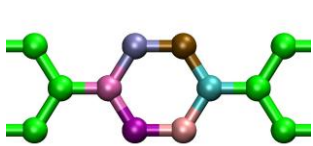
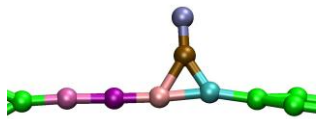
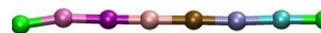

Figure 5-(a)　　　　　　　　Figure 5-(b)　　　　　　　　Figure 5-(c)

Figure 5 (best viewed in color) Elongation process from initial AGNR in (a) to intermediate configuration in (b) to MACC in (c)

In contrast, the ZGNR underwent a significantly different elongation process, as observed in Figure 6. There, we first saw the buckling of the initially six-atom hexagon in Figure 6(a) to the elongated rectangle in Figure 6(b), as previously discussed. Under continued tensile loading, the elongated rectangles in Figure 6(b) began to form a MACC as shown in Figure 6(c), with a very unusual deformation mechanism. In particular, it was observed that the rectangle joined the MACC by rotating the bond that connects the two parallel lines of carbon atoms in Figure 6(b), which formed a triangle-pentagon structure behind the MACC, as seen in Figure 6(c). Upon further tensile loading, the bond in the triangle that was nearly orthogonal to the tensile loading direction (which connects the brown and light blue atoms in Figure 6(c)) rotated in the direction of the tensile loading, and joined the MACC as seen in Figure 6(d), leaving a triangular structure in front of the elongated rectangles. Interestingly, subsequent deformation processes demonstrated, as seen in Figure 6(e) and 6(f), that the triangle-pentagon deformation mechanism remained operative, and controlled the subsequent elongation of the remainder of the non-MACC portion of the ZGNR into the MACC. Specifically, the transition between Figure 6(d) and 6(e) illustrate that the pentagon formation is energetically favorable as it is not possible to pull both bonds of the triangle in Figure 6(d) into the MACC simultaneously; as one of the triangular bonds is pulled into the MACC in Figure 6(e), the other dangling bond from the triangle in Figure 6(d) forms a pentagon with the rectangle directly connected to it in Figure 6(e) to minimize the system potential energy during the bond breaking process.

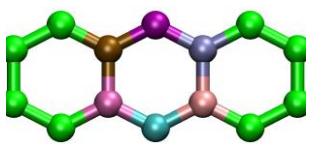
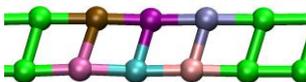
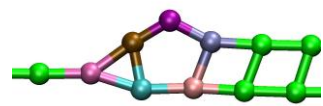

Figure 6(a)　　　　　　　　Figure 6(b)　　　　　　　　Figure 6(c)

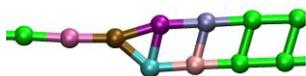 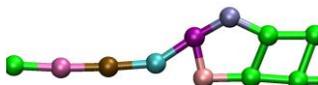 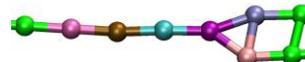

Figure 6(d)  Figure 6(e)  Figure 6(f)

Figure 6 Illustration of MACC formation through the triangle-pentagon structure under tensile loading for the ZGNR (best viewed in color; note the colors represent the same atoms of the AGNR in Figure 5 as for the ZGNR in this figure).

Our atomistic simulation results clearly show that full MACCs that consist of only single carbon atoms can be obtained from the tensile deformation of the narrowest (single unit cell wide) AGNR and ZGNR. We emphasize that MACC formation has not been observed in previous atomistic simulations of the tensile deformation of wider GNRs before fracture [21, 22, 30, 31]. In contrast, in the MD simulations of Wang et al. [16,17], a MACC was obtained from a larger GNR; however, they utilized an idealized initial configuration in which a single atom contact bridging two pieces of a wider GNR was used as a seed to form the MACC. Furthermore, upon elongation, only a small portion of the wider GNR was drawn into the MACC.

Our results are also consistent with the experimental studies of MACC formation of Jin et al. [11], who used controlled electron irradiation to thin GNRs to eventually form MACCs. Furthermore, it was noted in the work of Jin et al. [11] that MACC formation should be possible from both AGNR and ZGNR, which has been shown in our work. Furthermore, they calculated that the formation energy for MACCs decreases as the width of the GNR decreases; this is also consistent both with the previous atomistic simulations [21,22,30,31], where MACC formation was not found from the tensile deformation of wider GNRs, but also with our current work. Specifically, we performed MD simulations of the tensile elongation of 2-AGNR (two unit cells wide) and 2-ZGNR; the 2-ZGNR elongated to a MACC, while the 2-AGNR fractured at a strain of 50%, where no MACC formation was observed. All of these results taken together strongly suggest that a combination of GNR chirality and width is critical for the formation of full MACCs.

## 4. Conclusions

Our atomistic simulations have shown that elongated monatomic chains of carbon atoms can be pulled from ultranarrow GNR. More importantly, we have demonstrated that the chirality of the GNR has a significant influence on both the deformation mechanisms that are observed during the tensile loading, and the failure properties of the resulting monatomic carbon chain, as the hexagon rings of the nanoribbons were shown to exhibit distinct chirality-dependent deformation patterns and fracture characteristics. Finally, in comparison with previous atomistic studies on monatomic carbon chain formation [11, 16, 17] we have shown the utility of using ultranarrow (i.e. single unit cell wide) GNR to achieve elongated and stable MACCs of atoms with fracture strains exceeding 100%. Due to the current ability to experimentally synthesize narrow GNR with tailored edge functionality [32], the present results suggest that researchers may be able to exploit their distinct properties under tensile loading to create unique and novel components for future M/NEMS.

**Acknowledgements**

This research was supported by the National Basic Research Program of China under the Grant No. 2006CB300404 and NSF China under Grant No. 10902107.